\documentclass[twocolumn,aps,prd,preprintnumbers,showpacs,superscriptaddress,nofootinbib,amsmath,amssymb,floats,floatfix,showkeys,notitlepage,longbibliography]{revtex4-2}

\usepackage{orcidlink}
\usepackage[T1]{fontenc}
\usepackage{lmodern}
\usepackage{microtype}
\usepackage{amsmath,amssymb,bm}
\usepackage{graphicx}
\usepackage{booktabs}
\usepackage{array}
\usepackage{xcolor}
\newcommand{\dd}{\mathrm{d}}
\newcommand{\Mcal}{\mathcal{M}}
\newcommand{\Pcal}{\mathcal{P}}
\newcommand{\rhoh}{\widehat{\rho}}
\newcommand{\Msun}{M_{\odot}}
\newcommand{\sgr}{\mathrm{Sgr\,A}^{*}}
\newcommand{\mseven}{\mathrm{M87}^{*}}
\newcommand{\order}{\mathcal{O}}

\begin{document} \sloppy

\title{Density-preserving core-Einasto black holes with Event Horizon Telescope bounds}

\author{Ali \"Ovg\"un \orcidlink{0000-0002-9889-342X}}
\email{ali.ovgun@emu.edu.tr}
\affiliation{Physics Department, Eastern Mediterranean University, Famagusta, 99628 North Cyprus via Mersin 10, T\"urkiye}

\author{Reggie C. Pantig \orcidlink{0000-0002-3101-8591}} 
\email{rcpantig@mapua.edu.ph}
\affiliation{Physics Department, School of Foundational Studies and Education, Map\'ua University, 658 Muralla St., Intramuros, Manila 1002  Philippines.}

\begin{abstract}
A Newtonian halo density does not uniquely determine a relativistic spacetime,
and the resulting completion ambiguity can dominate predicted horizon-scale
signals. We make this dependence explicit for the feedback-cored Einasto
profile calibrated on FIRE-2 simulations.  In Schwarzschild gauge, the Einstein
equations give an asymptotically flat, density-preserving geometry supported by
an effective anisotropic fluid; no microscopic description of collisionless
dark matter is assumed.  By contrast, a commonly used rotation-curve
completion replaces the finite seed core by an inverse-square source cusp.
For the density-preserving solution we derive the enclosed mass in closed form,
obtain a sharp one-horizon criterion and the extremal boundaries of a
three-horizon phase, and establish the fixed-halo first law.  More generally,
the leading fractional shift of a spherical black-hole shadow is the
environmental mass enclosed within the vacuum photon sphere divided by the
central black-hole mass.  Applying this result to published Event Horizon
Telescope shadow-deviation summaries gives illustrative one-sided 95\%
credible limits of $4.8\times10^{23}\,\Msun\,{\rm pc}^{-3}$ for $\sgr$ and
$4.1\times10^{17}\,\Msun\,{\rm pc}^{-3}$ for $\mseven$, far above realistic
smooth-halo densities.  A Milky-Way calibration predicts a fractional shadow
shift of $1.7\times10^{-25}$, whereas the rotation-curve completion gives a
shift 19 orders of magnitude larger for the same galactic inputs.  A covariant
polarized thin-disk calculation shows the same weak-core suppression. Thus
current horizon-scale images do not constrain a smooth kiloparsec-scale
core-Einasto halo; an observable environmental signal would instead require a
compact inner component or a physically different relativistic source.
\end{abstract}

\maketitle

\section{Introduction}
\label{sec:introduction}

The exterior of an astrophysical black hole is not an exact vacuum. Accreting plasma, stars, and dark matter can perturb particle and photon trajectories and thereby modify horizon-scale observables \citep{Cardoso2022,Figueiredo2023}. Dark-matter spikes may develop through the adiabatic growth of a central black hole \citep{GondoloSilk1999}, while the surrounding galactic halo is commonly described by empirical profiles such as Navarro--Frenk--White, Burkert, and Einasto \citep{NFW1997,Burkert1995,Einasto1965}. The advent of horizon-scale imaging therefore raises a basic modeling question: which part of a predicted image displacement follows from the assumed density profile, and which part is introduced by the relativistic completion used to embed that profile around a black hole?

A density law alone cannot determine the answer. It specifies only one function, whereas a static, spherically symmetric metric and its supporting anisotropic stress tensor contain additional functional freedom. One widely used prescription integrates a Newtonian rotation curve to construct the temporal lapse, imposes $g_{tt}g_{rr}=-1$, and then adds a Schwarzschild term \citep{Xu2018,Hou2018}. Direct source-level analyses have shown, however, that the resulting Einstein tensor need not reproduce the density profile from which the construction began \citep{Datta2024,Bolokhov2026}. This observation has motivated density-preserving relativistic completions of several standard halo profiles \citep{KonoplyaZhidenko2022,Figueiredo2023}. It is especially relevant to observational inference because a recent EHT analysis of an ordinary Einasto halo employed an interpolating completion to constrain halo parameters \citep{Errehymy:2026wti}. Any such constraint is necessarily conditional on the metric closure---and hence on the effective stress tensor---adopted in the analysis.

Relativistic studies of dark-matter environments now cover a broad range of matter models and observables. Black-hole geometries sourced by quantum-wave solitons and by the Dekel--Zhao profile have been used to study shadows, lensing, and accretion signatures \citep{Pantig:2022sjb,Ovgun:2025bol}, whereas finite-distance weak deflection, gravitational ringing, and superradiant quasibound states probe complementary radial and dynamical regimes \citep{Pantig:2022toh,Liu:2022ygf}. Perfect-fluid dark matter has also been coupled to Euler--Heisenberg nonlinear electrodynamics. In these backgrounds, periodic-orbit and numerical-kludge waveforms separate environmental effects from quantum-electrodynamic corrections \citep{Gogoi:2026obd}, while scattering amplitudes, greybody factors, Hawking emission, and neutrino energy deposition provide wave and thermal diagnostics \citep{Becar:2026doz}. Beyond black holes, dark-matter profiles can support evolving, topologically deformed wormholes \citep{Ovgun:2018uin}, illustrating that the choice of matter model can alter the global compact-object geometry rather than merely perturb a vacuum solution.

The geometric basis of strong-field imaging is supplied by photon surfaces and unstable photon orbits, which control relativistic images and the critical curve associated with a black-hole shadow \citep{Claudel:2000yi,Virbhadra:2007kw,Virbhadra:2008ws,Virbhadra:2024xpk,Perlick:2015vta,Kobialko:2025sls}. This framework has been extended to higher-dimensional and other non-Kerr black holes \citep{Atamurotov:2013sca,Papnoi:2014aaa}, rotating regular solutions and compact-object mimickers \citep{Abdujabbarov:2016hnw,Abdikamalov:2019ztb}, and plasma-dressed geometries \citep{Atamurotov:2015nra,Ovgun:2026vcl}. Further examples include four-dimensional Einstein--Gauss--Bonnet and quantum-effective Schwarzschild black holes, as well as accelerating, Lorentzian--Euclidean, quasi-topological, and scalarized solutions \citep{Kumar:2020owy,Wang:2025fmz,Chakhchi:2024obi,Battista:2026nsx,Lutfuoglu:2026gis,Lutfuoglu:2026gey}. Shadow observables have also been cast as parameter-estimation tools and applied to EHT constraints on rotating hairy spacetimes \citep{Kumar:2018ple,Afrin:2021imp}. These developments make clear that a shift of the metric critical curve must be distinguished from a change in the observed brightness depression, which also depends on the emitting matter and photon propagation.

Horizon-scale polarimetry provides an independent probe of this distinction. Its covariant foundation lies in relativistic radiative-transfer theory \citep{Lindquist:1966igj}, which is now implemented in public general-relativistic polarized-transfer codes \citep{Dexter:2016cdk,Huang:2024bar}. Simplified equatorial synchrotron models show explicitly how the magnetic-field geometry, orbital motion, strong lensing, and parallel transport combine to produce the observed electric-vector position-angle pattern \citep{Narayan2021Polarized,Gelles2021Polarized}. Related calculations have explored rotating scalar--tensor--vector-gravity black holes, magnetized Kerr jet trajectories, Bonnor diholes, and traversable wormholes \citep{Qin:2022kaf,Zhang:2023cuw,Zhang:2022klr,Delijski:2022jjj}. Circular polarization can further diagnose magnetic-field handedness, while general-relativistic magnetohydrodynamic simulations identify reconnection and plasmoid formation as possible sources of time-dependent near-horizon structure \citep{Ricarte:2021frd,Nathanail:2021jbn}. The distinction between the photon ring and the emission-dependent inner shadow \citep{Chael:2021rjo}, together with QED-induced modifications of photon propagation \citep{Hu:2020usx}, reinforces the need to separate spacetime geometry from transfer physics.

The polarized EHT images of $\mseven$ reveal an ordered ring and favor dynamically important magnetic fields near the horizon \citep{EHT2021Polarization,EHT2021Magnetic}. Environmental imaging has recently been investigated in perfect-fluid backgrounds \citep{Huang:2026cjn}, and an exact Schwarzschild--Hernquist halo was found to produce sub-percent changes in the direct image for galaxy-compatible parameters \citep{Angelov2025}. These results motivate asking whether the suppression of geometric shadow shifts in a smooth central core also persists in polarization-sensitive observables.

In this work, we prescribe the \emph{core-Einasto} (cEinasto) profile introduced by Lazar \emph{et al.} to describe baryonic-feedback cores in 54 FIRE-2 haloes \citep{Lazar2020}. Its shifted-radius dependence,
\begin{equation}
\rho_{\rm cE}(r)\propto
\exp\left[-\frac{2}{\alpha}
\left(\frac{r+r_c}{r_s}\right)^{\alpha}\right],
\end{equation}
distinguishes it both from the ordinary Einasto law and from regular-black-hole sources sometimes described as Einasto cores \citep{Alshammari2025,KonoplyaZhidenko2026}. To the best of our knowledge, the cEinasto law has not previously been imposed as the prescribed density of a black-hole spacetime. We close the Einstein equations by adopting the Schwarzschild gauge $g_{tt}g_{rr}=-1$. The radial and tangential pressures then follow from the field equations rather than from a microscopic dark-matter model. Accordingly, the resulting matter sector is interpreted as an effective anisotropic source, not as a kinetic description of cold, collisionless dark matter.

Our central result is a pronounced hierarchy of scales. For any weak, spherically symmetric density profile treated within this closure, the leading shift of the shadow scale is governed by the environmental mass enclosed by the vacuum photon sphere. A smooth cEinasto core therefore produces a fractional effect of order $\rho_0M_\bullet^2$. By contrast, the rotation-curve completion can generate an order-$\rho_sr_s^2$ redshift through its lapse normalization. For the Milky-Way benchmark parameters used in this work, the two prescriptions differ by approximately 19 orders of magnitude. This discrepancy is not a numerical subtlety; it reflects the inequivalent effective stress tensors encoded by the two relativistic completions.

Our contributions are fivefold. First, we derive the cEinasto mass function and gravitational potential in analytic form and identify explicitly the source mismatch produced by the rotation-curve completion. Second, we construct an exact density-preserving black-hole geometry and determine its horizon phase structure and fixed-halo thermodynamics. Third, we obtain a compact weak-environment shadow formula valid for any smooth spherical density profile within the adopted closure. Fourth, we translate the published EHT ring-diameter constraints for $\sgr$ and $\mseven$ into transparent, summary-level bounds. Fifth, we compute polarized synchrotron images using covariant ray tracing and isolate the weak-core scaling of image-domain observables. The polarized thin-disk calculation is intended as a controlled geometric test, not as a substitute for polarized general-relativistic magnetohydrodynamic inference. Unless stated otherwise, we use $G=c=\hbar=k_B=1$ and the metric signature $(-,+,+,+)$; physical units are restored in Sec.~\ref{sec_eht}.

\section{Why a density profile does not fix a metric}
\label{sec_audit}

Consider the Schwarzschild-gauge line element
\begin{equation}
 \dd s^2=-f(r)\dd t^2+\frac{\dd r^2}{f(r)}
 +r^2(\dd\theta^2+\sin^2\theta\,\dd\phi^2).
 \label{eq_metric}
\end{equation}
For an anisotropic stress tensor
$T^{\mu}{}_{\nu}=\operatorname{diag}(-\rho,p_r,p_t,p_t)$, the independent
Einstein equations give
\begin{align}
 8\pi\rho&=\frac{1-f-rf'}{r^2},                                      \label{eq_einsteinrho}\\
 p_r&=-\rho, \qquad
 p_t=-\rho-\frac{r}{2}\rho'.                                       \label{eq_pressures}
\end{align}
Thus the gauge choice $g_{tt}g_{rr}=-1$ is already an equation-of-state
closure.  It cannot simultaneously be imposed as a harmless metric
simplification for pressureless matter.

For a circular timelike geodesic, the physical tangential speed measured by a
static observer obeys the exact identity
\begin{equation}
 v_c^2=\frac{r f'}{2f}.                                               \label{eq_vcgr}
\end{equation}
The rotation-curve prescription equates Eq. \eqref{eq_vcgr} to the Newtonian
quantity $M_h(r)/r$ and integrates it.  If the Newtonian potential is denoted
by $\Phi_N$ and normalized to vanish at infinity, the resulting halo lapse is
$H=\exp(2\Phi_N)$.  Appending a central mass gives
\begin{equation}
 f_{\rm RC}(r)=H(r)-\frac{2M_\bullet}{r}.                             \label{eq_frc}
\end{equation}
Substitution into Eq. \eqref{eq_einsteinrho} yields
\begin{equation}
 8\pi\rho_{\rm RC}=\frac{1-H-rH'}{r^2}.                              \label{eq_rhoRCgeneral}
\end{equation}
The $M_\bullet$ contribution cancels identically.  Any difference
between an advertised Newtonian seed density and $\rho_{\rm RC}$ must therefore
arise from the chosen relativistic prescription rather than from a
black-hole-induced cusp.

This observation directly revises the interpretation of the central cusps
reported using the prescription of Ref. \cite{Xu2018}.  It also
cautions against applying a Newman--Janis transformation unless we first
verify the static source.  Rotating an inconsistent seed does not restore the
intended matter profile. We consequently restrict our analysis
to the exact static system, for which every component of
$T^{\mu}{}_{\nu}$ is explicit.

\section{FIRE-2 core-Einasto profile}
\label{sec_profile}

Introduce dimensionless variables
\begin{equation}
 x=\frac{r}{r_s},\qquad q=\frac{r_c}{r_s},\qquad k=\frac{2}{\alpha}.
\end{equation}
The cEinasto density is \cite{Lazar2020}
\begin{equation}
 \rho(r)=\rho_s\rhoh(x),\qquad
 \rhoh(x)=\exp\!\left\{-k\left[(x+q)^\alpha-1\right]\right\}.
 \label{eq_rhoce}
\end{equation}
It has the finite central density
$\rho_0=\rho_s\exp[k(1-q^\alpha)]$ and logarithmic slope
\begin{equation}
 \frac{\dd\ln\rho}{\dd\ln r}=-2x(x+q)^{\alpha-1}.                   \label{eq_slope}
\end{equation}
The FIRE-2 analysis found that fixing $\alpha=0.16$ describes a broad range of
feedback-affected haloes, with typical cores of $0.5$--$2\,\mathrm{kpc}$ in
Milky-Way-mass systems and $1$--$5\,\mathrm{kpc}$ in bright dwarfs
\cite{Lazar2020}.

Define
\begin{equation}
 M_h(r)=4\pi\rho_s r_s^3\Mcal(x),\qquad
 \Mcal(x)=\int_0^x y^2\rhoh(y)\,\dd y.                              \label{eq_Mdefinition}
\end{equation}
With $z_x=k(x+q)^\alpha$, $z_0=kq^\alpha$, and
$\Delta\gamma_a=\gamma(a,z_x)-\gamma(a,z_0)$, direct integration gives
\begin{align}
 \Mcal(x)=\frac{e^k}{\alpha}\big[&k^{-3/\alpha}\Delta\gamma_{3/\alpha}
 -2qk^{-2/\alpha}\Delta\gamma_{2/\alpha}\nonumber\\
 &+q^2k^{-1/\alpha}\Delta\gamma_{1/\alpha}\big].                 \label{eq_Manalytic}
\end{align}
This finite-mass expression is useful both numerically and analytically.  The
total halo mass follows by replacing each lower incomplete gamma difference
with the corresponding upper incomplete gamma function at $z_0$.

\begin{figure*}[t]
 \centering
 \includegraphics[width=0.92\textwidth]{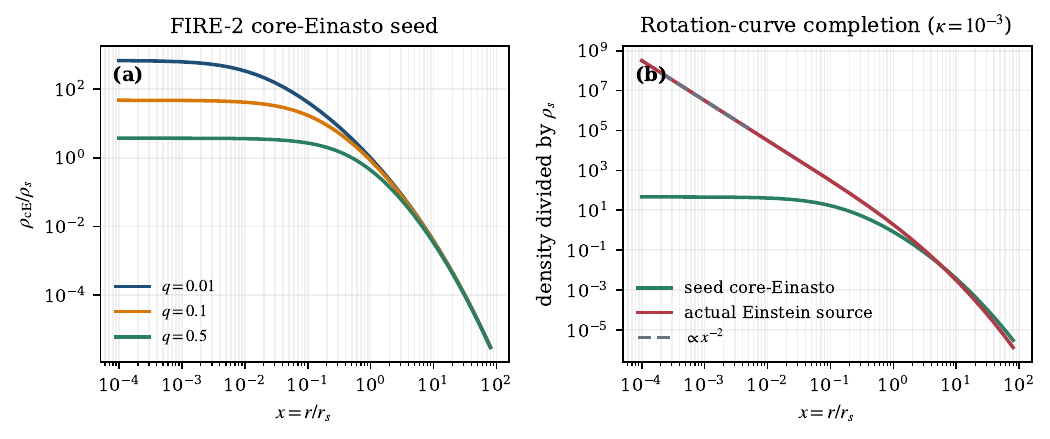}
 \caption{\label{fig:profileaudit}
 (a) FIRE-2 cEinasto density for three dimensionless core radii at fixed
 $\alpha=0.16$.  (b) The prescribed cEinasto density and the density obtained
 by inserting the rotation-curve lapse into the Einstein equations.  Despite
 the finite seed core, the latter develops an artificial $x^{-2}$ cusp.  The
 black-hole mass does not enter this comparison because its contribution to
 Eq. \eqref{eq_einsteinrho} cancels.}
\end{figure*}

The positive dimensionless potential function satisfying
$\Pcal(\infty)=0$ and $\Pcal'=-\Mcal/x^2$ is
\begin{align}
 \Pcal(x)={}&\frac{\Mcal(x)}{x}+\frac{e^k}{\alpha}
 \left[k^{-2/\alpha}\Gamma\!\left(\frac{2}{\alpha},z_x\right)\right.\nonumber\\
 &\left.\hspace{22mm}-qk^{-1/\alpha}
 \Gamma\!\left(\frac{1}{\alpha},z_x\right)\right].
 \label{eq_potential}
\end{align}
The Newtonian potential is $\Phi_N=-4\pi\rho_s r_s^2\Pcal$.  Hence the lapse
used by the rotation-curve prescription is
\begin{equation}
 H(x)=e^{-\kappa\Pcal(x)},\qquad
 \kappa=8\pi\rho_s r_s^2.                                         \label{eq_Hseed}
\end{equation}
Its actual source, in units of the seed normalization, is
\begin{equation}
 \frac{\rho_{\rm RC}}{\rho_s}=
 \frac{1-H(x)[1+\kappa\Mcal(x)/x]}{\kappa x^2}.                    \label{eq_rhoRC}
\end{equation}
Since $H_0=e^{-\kappa\Pcal(0)}<1$, Eq. \eqref{eq_rhoRC} behaves as
$(1-H_0)/(\kappa x^2)$ at the origin, although Eq. \eqref{eq_rhoce} is cored.
Figure \ref{fig:profileaudit} displays the mismatch.  This exact
$x^{-2}$ asymptotic behavior is the fingerprint of the completion,
not a response of the seed halo to $M_\bullet$.

\section{Exact density-preserving geometry}
\label{sec_solution}

The prescribed density is retained by defining the Misner--Sharp mass
\begin{equation}
 m(r)=M_\bullet+M_h(r),\qquad f(r)=1-\frac{2m(r)}{r}.
 \label{eq_massmetric}
\end{equation}
Equations \eqref{eq_einsteinrho}--\eqref{eq_pressures} are then satisfied
identically by Eq. \eqref{eq_rhoce} and 
\begin{equation}
 f(x)=1-\frac{2\lambda}{x}-\kappa\frac{\Mcal(x)}{x},\qquad
 \lambda=\frac{M_\bullet}{r_s}.      \label{eq_metricdimensionless}
\end{equation}
Because $\Mcal(\infty)$ is finite, the spacetime is asymptotically flat with
$M_{\rm ADM}=M_\bullet+4\pi\rho_s r_s^3\Mcal(\infty)$.  The
central Schwarzschild singularity remains when $M_\bullet>0$.  The halo itself
is regular and contributes a de Sitter-like term
\begin{equation}
 f(r)=1-\frac{2M_\bullet}{r}-\frac{8\pi\rho_0}{3}r^2+\order(r^3/r_c).
 \label{eq_coreexpansion}
\end{equation}

The source has a transparent energy-condition structure.  Since $\rho\ge0$
and $\rho'\le0$, the radial null energy condition is saturated,
$\rho+p_r=0$, while
\begin{equation}
 \rho+p_t=-\frac{r\rho'}{2}
 =\rho\,x(x+q)^{\alpha-1}\ge0.                                    \label{eq_NEC}
\end{equation}
The null and weak energy conditions therefore hold everywhere.  The strong
condition requires $p_t\ge0$ and fails in the inner core.  The
dominant condition further requires $x(x+q)^{\alpha-1}\le2$.  It can fail only
in the exponentially dilute outer tail.  These pressures are the price of the
one-function gauge closure.  The solution should accordingly be interpreted
as a dark-matter-inspired effective anisotropic source, not as a kinetic
solution for a cold collisionless distribution \cite{Datta2024}.

\section{Horizons and phase structure}
\label{sec_horizons}

Killing horizons are the positive roots of
\begin{equation}
 \mathcal H(x)=x-2\lambda-\kappa\Mcal(x)=0,                         \label{eq_horizoneq}
\end{equation}
whose derivative is
\begin{equation}
 \mathcal H'(x)=1-\kappa x^2\rhoh(x).                              \label{eq_horizonderivative}
\end{equation}
The function $x^2\rhoh(x)$ has one maximum.  Its location $x_*$ is the unique
positive solution of
\begin{equation}
 x_*(x_*+q)^{\alpha-1}=1.                                           \label{eq_xstar}
\end{equation}
It follows that
\begin{equation}
 \kappa_{\rm crit}=\frac{1}{x_*^2\rhoh(x_*)}                        \label{eq_kcrit}
\end{equation}
is a sharp monotonicity threshold.  For
$0\le\kappa\le\kappa_{\rm crit}$, $\mathcal H$ rises from
$-2\lambda$ to infinity and there is exactly one horizon.  Above
this threshold two extrema appear.  Depending on $\lambda$, the geometry has
one, two degenerate, or three Killing horizons.  The extremal boundaries admit the
parametric representation
\begin{equation}
 \kappa_e(x_e)=\frac{1}{x_e^2\rhoh(x_e)},\qquad
 \lambda_e(x_e)=\frac{x_e-\kappa_e\Mcal(x_e)}{2}.                   \label{eq_extremal}
\end{equation}

\begin{figure*}[t]
 \centering
 \includegraphics[width=0.94\textwidth]{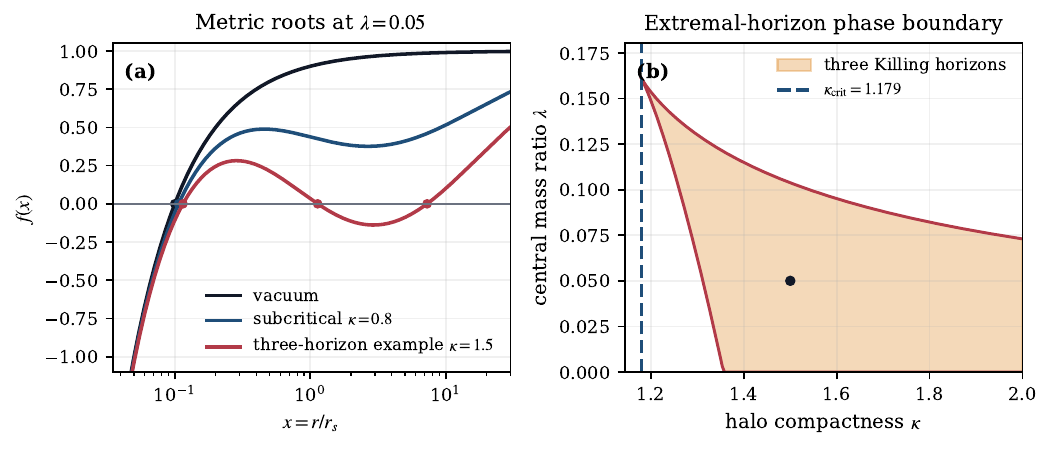}
 \caption{\label{fig:horizons}
 (a) Metric function for a vacuum black hole, a subcritical cEinasto halo, and
 a strong-halo configuration with three roots.  The logarithmic radial scale
 resolves the inner pair.  (b) The extremal curves follow from
 Eq. \eqref{eq_extremal}.  The shaded domain contains three Killing horizons.
 The black point marks the example in panel (a).}
\end{figure*}

For the FIRE-2 shape $\alpha=0.16$ and $q=0.1$, we obtain
$\kappa_{\rm crit}=1.179145105$.  Figure \ref{fig:horizons} shows a
three-horizon example at $(\kappa,\lambda)=(1.5,0.05)$ and the complete
extremal boundary in this slice of shape space.  Galactic values are typically
$\kappa\lesssim10^{-5}$ and therefore lie safely in the unique-horizon phase.
The multi-horizon sector is a mathematically consistent strong-source limit,
not a regime supported by the FIRE-2 calibration.

\section{Fixed-halo thermodynamics}
\label{sec_thermo}

Let $x_h$ be a simple outer horizon.  From Eq. \eqref{eq_horizonderivative},
the Hawking temperature and Bekenstein--Hawking entropy are
\begin{align}
 T_H&=\frac{1-\kappa x_h^2\rhoh(x_h)}{4\pi r_sx_h},                 \label{eq_temperature}\\
 S&=\pi r_s^2x_h^2.                                                \label{eq_entropy}
\end{align}
At fixed $(\rho_s,r_s,r_c,\alpha)$, the horizon equation implies
\begin{equation}
 \frac{\dd\lambda}{\dd x_h}=\frac{1-\kappa x_h^2\rhoh(x_h)}{2},
 \qquad \dd M_\bullet=T_H\,\dd S,                                 \label{eq_firstlaw}
\end{equation}
so the ordinary first law is exact along the one-parameter central-mass
family.  Varying halo parameters would introduce additional work terms and is
not part of this fixed-environment ensemble.

\begin{figure*}[t]
 \centering
 \includegraphics[width=0.77\textwidth]{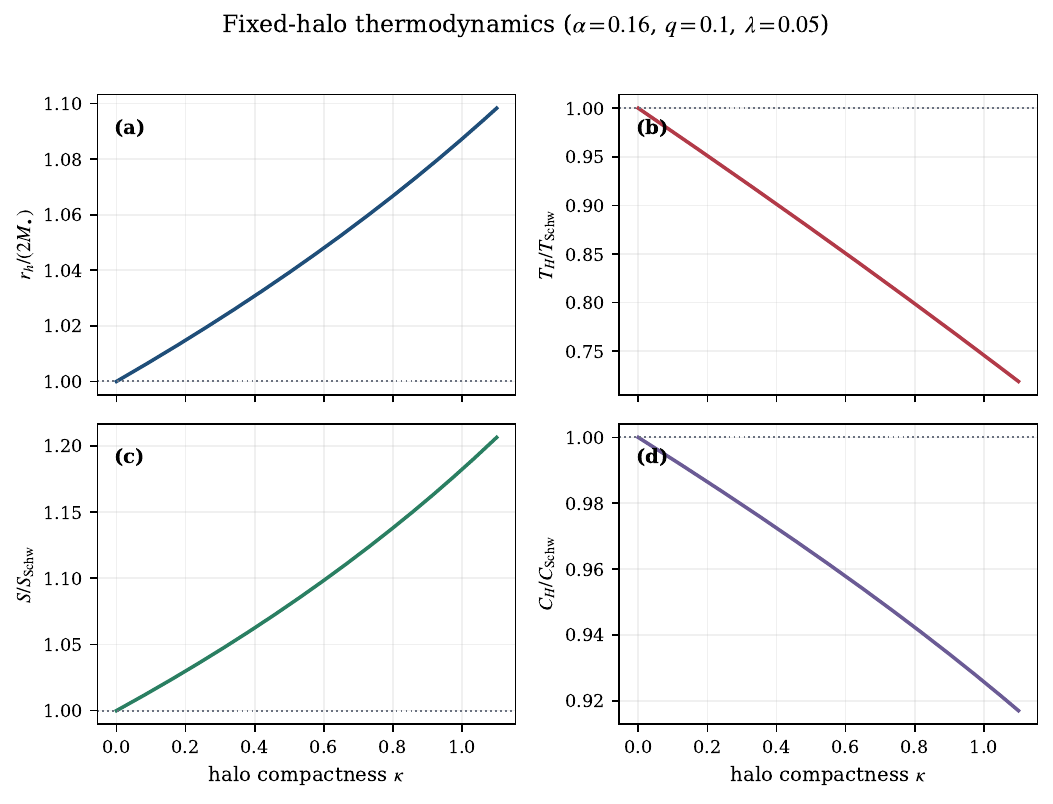}
 \caption{\label{fig:thermo}
 Exact fixed-halo thermodynamic observables relative to Schwarzschild values
 for $(\alpha,q,\lambda)=(0.16,0.1,0.05)$.  The halo enlarges the horizon and
 entropy and lowers the temperature.  The heat capacity remains on the
 negative Schwarzschild-connected branch throughout the displayed
 subcritical interval.}
\end{figure*}

The corresponding heat capacity is
\begin{equation}
 C_H=-2\pi r_s^2x_h^2
 \frac{1-\kappa x_h^2\rhoh(x_h)}
 {1+\kappa x_h^2[\rhoh(x_h)+x_h\rhoh'(x_h)]}.                      \label{eq_heatcapacity}
\end{equation}
The temperature vanishes on an extremal boundary.  Its stationary points,
and hence possible heat-capacity divergences, occur when the denominator of
Eq. \eqref{eq_heatcapacity} vanishes.  In the astrophysical weak-halo regime,
$C_H=-8\pi M_\bullet^2[1+\order(\kappa)]<0$ and no thermodynamically stable
branch is generated.  Figure \ref{fig:thermo} follows the exact quantities to
large compactness while remaining below $\kappa_{\rm crit}$.

For a photon sphere well inside the smooth core it is useful to define
\begin{equation}
 \Lambda_0=8\pi\rho_0,\qquad u=\Lambda_0M_\bullet^2.
\end{equation}
Equations \eqref{eq_coreexpansion}--\eqref{eq_entropy} then yield
\begin{align}
 \frac{r_h}{2M_\bullet}&=1+\frac{4u}{3}+\order(u^2), &
 \frac{T_H}{T_{\rm Schw}}&=1-\frac{16u}{3}+\order(u^2),\nonumber\\
 \frac{S}{S_{\rm Schw}}&=1+\frac{8u}{3}+\order(u^2).              \label{eq_weakthermo}
\end{align}

\section{Photon sphere and shadow}
\label{sec_shadow}

For Eq. \eqref{eq_metric}, the null effective potential is proportional to
$f(r)/r^2$.  An unstable circular photon orbit therefore satisfies
$rf'-2f=0$ \cite{Claudel:2000yi,PerlickTsupko2022}.  In dimensionless variables this becomes
\begin{equation}
 x_{\rm ph}\left[1+\frac{\kappa}{2}x_{\rm ph}^2\rhoh(x_{\rm ph})\right]
 =3\left[\lambda+\frac{\kappa}{2}\Mcal(x_{\rm ph})\right],         \label{eq_photon}
\end{equation}
and the critical impact parameter seen from infinity is
\begin{equation}
 b_{\rm sh}=r_s\frac{x_{\rm ph}}{\sqrt{f(x_{\rm ph})}}.            \label{eq_shadowb}
\end{equation}
The relevant root is an exterior maximum of $f/r^2$.

A useful model-independent result follows by writing
$f=1-2[M_\bullet+M_h(r)]/r$ and treating $M_h/M_\bullet$ as first order.  Around
the Schwarzschild orbit,
\begin{align}
 r_{\rm ph}={}&3M_\bullet+3M_h(3M_\bullet)
 -3M_\bullet M_h'(3M_\bullet)+\order(M_h^2),                       \label{eq_rphpert}\\
 \delta_{\rm sh}\equiv{}&\frac{b_{\rm sh}}{3\sqrt3M_\bullet}-1
 =\frac{M_h(3M_\bullet)}{M_\bullet}+\order(M_h^2).                 \label{eq_deltageneral}
\end{align}
The derivative terms cancel from the observable shadow shift.  Smooth matter
affects the critical curve only through the mass enclosed at the unperturbed
photon sphere.  For a constant core, $M_h=4\pi\rho_0r^3/3$, giving
\begin{align}
 r_{\rm ph}&=3M_\bullet+\order(u^2),\nonumber\\
 \frac{b_{\rm sh}}{3\sqrt3M_\bullet}&=(1-9u)^{-1/2}
 =1+\frac{9u}{2}+\order(u^2).                                     \label{eq_desittershadow}
\end{align}
The cancellation in $r_{\rm ph}$ is the familiar Schwarzschild--de Sitter
result, while the critical impact parameter retains the redshift correction.

\begin{figure*}[t]
 \centering
 \includegraphics[width=0.94\textwidth]{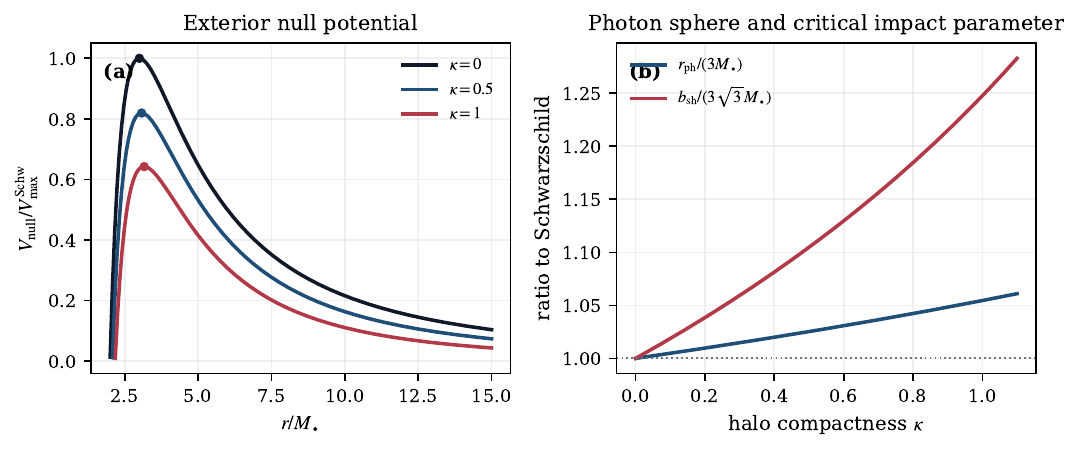}
 \caption{\label{fig:shadow}
 (a) We show the exterior null effective potential for several
 halo compactnesses at $(\alpha,q,\lambda)=(0.16,0.1,0.05)$.  Markers identify
 unstable photon orbits.  (b) Exact photon-sphere radius and critical impact parameter,
 normalized to their Schwarzschild values.  The shadow size is more sensitive
 than the coordinate photon-sphere radius.}
\end{figure*}

Figure \ref{fig:shadow} solves Eqs. \eqref{eq_photon} and
\eqref{eq_shadowb} without expansion.  Increasing $\kappa$ lowers and shifts
the null-potential maximum and enlarges the critical impact parameter.
These order-unity examples illustrate the geometry.  We address
realistic values next.

\section{EHT constraints and astrophysical scale}
\label{sec_eht}

The EHT measured a $51.8\pm2.3\,\mu$as emission-ring diameter for $\sgr$
\cite{EHTSgrI}.  After calibration against relativistic accretion-flow
simulations and combination with stellar-orbit priors, the inferred fractional
deviation of the shadow size from the Schwarzschild value is
\begin{equation}
 \delta=-0.08\pm0.09\quad({\rm VLTI}),\qquad
 \delta=-0.04^{+0.09}_{-0.10}\quad({\rm Keck})                     \label{eq_sgrdata}
\end{equation}
\cite{EHTSgrVI}.  For $\mseven$, the observed ring diameter is
$42\pm3\,\mu$as and the inferred gravitational angular scale is
$3.8\pm0.4\,\mu$as \cite{EHTM87VI}.  With the stellar-dynamical mass prior,
the commonly used shadow-deviation summary is $\delta=-0.01\pm0.17$
\cite{Psaltis2020}.

We translate these published summaries into an illustrative one-sided bound.
Each likelihood is approximated by a Gaussian in $\delta$, with the Keck
uncertainty symmetrized to $0.095$.  Because the positive-density static model
predicts $\delta\ge0$ at leading order, we impose a flat prior for
$\delta\ge0$ and zero support below it.  If $\Phi$ is the unit-normal CDF, the
$p$ credible upper endpoint is
\begin{equation}
 \delta_p=\mu+\sigma\Phi^{-1}\!\left[\Phi(-\mu/\sigma)
 +p\{1-\Phi(-\mu/\sigma)\}\right].                                \label{eq_truncnormal}
\end{equation}
We keep this inference deliberately modest.  We do not reanalyze
EHT visibilities or propagate the full calibration, spin, inclination, and
accretion-model posterior.

Inside a core much larger than the gravitational radius, we can
invert Eq. \eqref{eq_desittershadow} exactly at this order in the radial
expansion.
\begin{equation}
 u=\frac{1-(1+\delta)^{-2}}{9},\qquad
 \rho_0=\frac{u}{8\pi(G/c^2)(GM_\bullet/c^2)^2}.                    \label{eq_rhobound}
\end{equation}
Table \ref{tab_eht} lists the resulting bounds.  The VLTI-based $\sgr$ value
is the tighter Galactic-center constraint,
$\delta<0.132$ and
$\rho_0<4.79\times10^{23}\,\Msun\,\mathrm{pc}^{-3}$ at $95\%$
credibility.  The corresponding $\mseven$ limit is
$4.12\times10^{17}\,\Msun\,\mathrm{pc}^{-3}$.  Both are extraordinarily weak
as constraints on smooth galactic dark matter.

\begin{table*}[t]
\centering
\caption{\label{tab_eht}Positive-deviation summary constraints.  Density
limits use Eq. \eqref{eq_rhobound} and the mass associated with each published
shadow summary.}
\begin{tabular}{lcccc}
\toprule
Observation summary & $\delta_{68}$ & $\delta_{95}$ &
$\rho_{0,68}\,[\Msun\,{\rm pc}^{-3}]$ &
$\rho_{0,95}\,[\Msun\,{\rm pc}^{-3}]$\\
\midrule
$\sgr$ (VLTI, $4.297\times10^6\Msun$) & 0.0600 & 0.1316 & $2.41\times10^{23}$ & $4.79\times10^{23}$\\
$\sgr$ (Keck, $3.951\times10^6\Msun$) & 0.0776 & 0.1618 & $3.59\times10^{23}$ & $6.70\times10^{23}$\\
$\mseven$ ($6.5\times10^9\Msun$)       & 0.1644 & 0.3267 & $2.51\times10^{17}$ & $4.12\times10^{17}$\\
\bottomrule
\end{tabular}
\end{table*}

To display the scale separation, we choose a Milky-Way cEinasto profile with
$\alpha=0.16$, $q=0.05$, total halo mass $10^{12}\Msun$, and
$\rho(8.2\,\mathrm{kpc})=0.008\,\Msun\,\mathrm{pc}^{-3}$.  Solving these two
normalization conditions gives
\begin{align}
 r_s&=9.645\,\mathrm{kpc},\nonumber\\
 \rho_s&=6.49\times10^{-3}\,\Msun\,\mathrm{pc}^{-3},\nonumber\\
 \rho_0&=0.758\,\Msun\,\mathrm{pc}^{-3}.                           \label{eq_mwfid}
\end{align}
For the VLTI mass of $\sgr$, this corresponds to
\begin{align}
 \lambda&=2.13\times10^{-11},\qquad
 \kappa=7.27\times10^{-7},\nonumber\\
 \delta_{\rm sh}&=1.73\times10^{-25}.                              \label{eq_mwprediction}
\end{align}
The profile is therefore invisible to current shadow-size measurements.

For comparison only, an $\mseven$ scale estimate with
$M_h=10^{14}\Msun$, $r_s=100\,\mathrm{kpc}$, and the same $(\alpha,q)$ yields
$\rho_0=0.068\,\Msun\,\mathrm{pc}^{-3}$,
$\kappa=7.01\times10^{-6}$, and
$\delta_{\rm sh}=3.56\times10^{-20}$.  FIRE-2 did not calibrate the cEinasto
law for cluster-central ellipticals, so this number is an order-of-magnitude
illustration, not a fitted model of M87.

\begin{figure*}[t]
 \centering
 \includegraphics[width=0.94\textwidth]{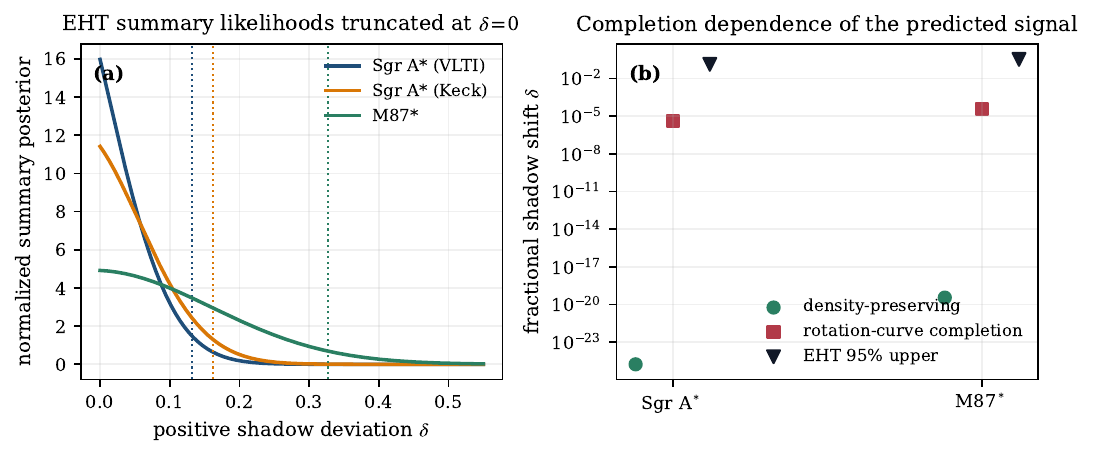}
 \caption{\label{fig:eht}
 (a) We show the Gaussian EHT shadow-deviation summaries after
 truncation to the positive-density model domain.  Vertical dotted lines mark
 the one-sided $95\%$ limits. (b) Fiducial density-preserving predictions, values from the
 rotation-curve completion, and EHT upper limits.  The vertical separation
 demonstrates both the weakness of current smooth-halo constraints and the
 physical importance of specifying the relativistic completion.}
\end{figure*}

The source audit explains why other completions can predict much larger
signals from the same galactic inputs.  Near the black hole the rotation-curve
lapse is approximately constant, $H\simeq H_0$.  Equations
\eqref{eq_frc} and \eqref{eq_shadowb} then give
\begin{equation}
 \delta_{\rm RC}=H_0^{-3/2}-1
 =\exp\!\left[\frac{3}{2}\kappa\Pcal(0)\right]-1,                 \label{eq_deltaRC}
\end{equation}
which is of order $\kappa$.  In contrast, the density-preserving solution has
\begin{equation}
 \delta_{\rm DP}=\frac{9}{2}\kappa\rhoh(0)\lambda^2
 +\order(\kappa\lambda^3/q,\kappa^2),                             \label{eq_deltaDP}
\end{equation}
which is suppressed by $\lambda^2$.  The Milky-Way values are
$\delta_{\rm RC}=3.94\times10^{-6}$ and
$\delta_{\rm DP}=1.73\times10^{-25}$.  Their
19-order-of-magnitude separation does not represent a numerical uncertainty.
It reflects inequivalent stress tensors.  Figure \ref{fig:eht} makes both the observational and completion
dependence explicit.

\section{Polarized synchrotron images}
\label{sec:polarization}

Linear polarimetry adds information that is absent from the total-intensity
shadow.  The observed electric-vector position angle (EVPA) depends jointly on
the magnetic field in the emitting plasma, relativistic motion, lensing, and
parallel transport along the photon trajectory
\citep{Narayan2021Polarized,Gelles2021Polarized}.  EHT observations of
$\mseven$ reveal a partially polarized ring with an ordered, approximately
azimuthal EVPA pattern, while comparisons with GRMHD calculations favor
magnetically arrested flows with dynamically important near-horizon fields
\citep{EHT2021Polarization,EHT2021Magnetic}.  Polarization is therefore a
sensitive probe, but not a metric-only observable \cite{Huang:2026cjn}.

A recent calculation for a Schwarzschild black hole in an exact Hernquist halo
found sub-percent changes in direct polarized images for galaxy-compatible
halo parameters and changes below ten percent even for indirect images at low
inclination \citep{Angelov2025}.  Here we perform the corresponding calculation
for the density-preserving cEinasto geometry.  Because the cEinasto spacetime is static
that Eq.~\eqref{eq_metricdimensionless} is static.  We consequently do not import the
spin-dependent constants, disk kinematics, or frame dragging of a Kerr-like
perfect-fluid-dark-matter metric.  Instead, we solve the null and polarization
transport equations directly in the exact spacetime derived in
Sec.~\ref{sec_solution}.  This isolates the monopolar environmental effect and
does not presume that a Newman--Janis transform preserves the cEinasto source.

\subsection{Backward rays and circular-disk kinematics}
\label{sec:polgeodesics}

For a past-directed ray launched from the observer, choose the conserved
covariant energy $p_t=E>0$ and azimuthal angular momentum $p_\phi=L_z$.  The
null Hamiltonian in the metric \eqref{eq_metric} is
\begin{equation}
 2\mathcal{H}=-\frac{E^2}{f}+f p_r^2+\frac{p_\theta^2}{r^2}
 +\frac{L_z^2}{r^2\sin^2\theta}=0.                               \label{eq:polham}
\end{equation}
Hamilton's equations give
\begin{align}
 \dot r&=fp_r,&
 \dot\theta&=\frac{p_\theta}{r^2},                              \nonumber\\
 \dot\phi&=\frac{L_z}{r^2\sin^2\theta},                         \label{eq:polx}\\
 \dot p_r&=-\frac12\left(\frac{E^2f'}{f^2}+f'p_r^2
 -\frac{2p_\theta^2}{r^3}
 -\frac{2L_z^2}{r^3\sin^2\theta}\right),                       \label{eq:polpr}\\
 \dot p_\theta&=\frac{L_z^2\cos\theta}{r^2\sin^3\theta},      \label{eq:polpth}
\end{align}
where the overdot denotes differentiation with respect to an affine
parameter.  Spherical symmetry supplies the additional invariant
$L^2=p_\theta^2+L_z^2/\sin^2\theta$.  For an observer at inclination $i$ in
the asymptotic region, the screen coordinates are
\begin{equation}
 \alpha_{\rm sky}=-\frac{L_z}{E\sin i},\qquad
 \beta_{\rm sky}=\frac{p_\theta}{E}.                              \label{eq:polscreen}
\end{equation}

The emitting plasma follows prograde circular geodesics in the equatorial
plane.  Its angular velocity and four-velocity are
\begin{align}
 \Omega^2&=\frac{f'}{2r},\nonumber\\
 u^\mu&=u^t(1,0,0,\Omega),\qquad
 u^t=\left(f-r^2\Omega^2\right)^{-1/2}.                           \label{eq:polfluid}
\end{align}
The marginal-stability condition is
\begin{equation}
 3ff'+rff''-2r(f')^2=0,                                           \label{eq:isco}
\end{equation}
and the disk begins at the first stable root exterior to the unstable photon
orbit.  This inner boundary is important: continuing the Keplerian model to
the horizon would assign circular emitters to a region where the assumed
timelike orbits do not exist.

\subsection{Emission and covariant polarization transport}
\label{sec:polemission}

Let $\{e_{(a)}{}^\mu\}$ be the orthonormal tetrad comoving with the circular
fluid.  We prescribe the magnetic field by its spatial tetrad components
$B^{(i)}$ and construct the emitted electric-polarization direction as
\begin{equation}
 \mathfrak f^\mu
 =\mathcal N\,\epsilon^{\mu\nu\rho\sigma}u_\nu k_\rho B_\sigma,
 \qquad \mathfrak f\cdot u=\mathfrak f\cdot k=0,                  \label{eq:polvector}
\end{equation}
where $k^\mu$ is future directed and $\mathcal N$ fixes
$\mathfrak f^2=1$.  The optically thin synchrotron weight includes the pitch
angle
\begin{equation}
 \sin^2\zeta=\frac{|\bm{k}\times\bm{B}|^2}
 {|\bm{k}|^2|\bm{B}|^2}                                         \label{eq:pitch}
\end{equation}
in the fluid frame.  Faraday rotation, conversion, absorption, and
self-consistent magnetohydrodynamics are not included; the calculation is a
geometric emission model, not a replacement for polarized GRMHD transfer.

At the observer we initialize two orthonormal screen vectors
$s_A{}^\mu$ ($A=1,2$), each orthogonal to the ray, and evolve them backward by
\begin{equation}
 k^\nu\nabla_\nu s_A{}^\mu=0.                                    \label{eq:parallel}
\end{equation}
This is equivalent to transporting $\mathfrak f^\mu$ forward, but avoids a
separate initial-value problem for every disk crossing.  Polarization is an
equivalence class under $s_A{}^\mu\rightarrow s_A{}^\mu+c_Ak^\mu$; the code
removes this pure-gauge component in a local static tetrad and orthonormalizes
the two-dimensional quotient screen.  At emission,
$f_A=\mathfrak f_\mu s_A{}^\mu$ determines
\begin{equation}
 e^{2i\chi}=\frac{(f_1+if_2)^2}{f_1^2+f_2^2}.                    \label{eq:evpa}
\end{equation}

For each equatorial crossing, the redshift is
$g=(-k\cdot u_{\rm obs})/(-k\cdot u_{\rm em})$.  We adopt the radial
emissivity used in the reference equatorial-disk construction \cite{Huang:2026cjn},
\begin{equation}
 \ln J=-2\ln\!\left(\frac{r}{r_h}\right)
 -\frac12\ln^2\!\left(\frac{r}{r_h}\right)+\text{constant},     \label{eq:emissivity}
\end{equation}
with the constant chosen so that $J(r_{\rm ISCO})=1$.  The observed Stokes
parameters in one pixel are
\begin{align}
 I&=\sum_n g_n^3J_n,\nonumber\\
 Q+iU&=p_0\sum_n g_n^3J_n\sin^2\zeta_n\,e^{2i\chi_n},            \label{eq:stokes}
\end{align}
where $p_0=0.70$ is the intrinsic polarization fraction.  The first disk
crossing defines the direct image; the second and third crossings are grouped
as higher-order emission.

We use two image-domain diagnostics.  The net linear-polarization fraction and
the second azimuthal polarization mode are
\begin{align}
 |m|_{\rm net}&=\frac{|\sum_j(Q_j+iU_j)|}{\sum_j I_j},             \label{eq:mnet}\\
 \beta_2&=\frac{\sum_j(Q_j+iU_j)e^{-2i\varphi_j}}{\sum_j I_j},    \label{eq:beta2}
\end{align}
where $\varphi_j=\arg(\alpha_{{\rm sky},j}+i\beta_{{\rm sky},j})$.
The magnitude $|\beta_2|$ measures coherent rotational structure; its phase
tracks the handedness of the EVPA pattern.  An absolute phase comparison
requires the same sky parity and Stokes convention, so the robust metric
quantity here is the change relative to the $\kappa=0$ image.

\subsection{Numerical setup and validation}
\label{sec:polnumerics}

For the fiducial images, we use $(\alpha,q,\lambda)=(0.16,0.1,0.05)$,
$i=17^\circ$, $r_{\rm obs}=10^4M_\bullet$, a
$22M_\bullet\times22M_\bullet$ field of view sampled on a $120^2$ grid, and
an emitting disk from $r_{\rm ISCO}$ to $20M_\bullet$.  We choose the
fluid-frame field $B^{(i)}=(0.87,0.50,0)$, in the order
$(r,\theta,\phi)$, to display the effect of an ordered polar component.  The
Stokes maps are convolved with a 1.05-pixel Gaussian solely to suppress pixel
aliasing.  A uniform $84^2$ grid is used for the nine-point compactness scan.

The sequence $0\leq\kappa\leq0.2$ is an illustrative high-compactness sequence, not a fit to
a galactic halo.  It remains on the one-horizon branch and keeps the ISCO
inside the emitting domain.  Both exact spherical invariants are projected
after every Runge--Kutta step.  At contributing disk crossings, the largest
relative null-Hamiltonian residual is $5.6\times10^{-16}$, the largest screen
norm or orthogonality residual is $6.7\times10^{-16}$, and every ray
terminates.  The interpolated mass obeys
$\Mcal'(x)=x^2\rhoh(x)$ to a maximum relative error
$8.1\times10^{-4}$ on the interpolation grid.  Increasing the image resolution
from $120^2$ to $160^2$ changes $|m|_{\rm net}$ by $5.3\%$ and
$\arg\beta_2$ by $0.63^\circ$ at $\kappa=0.1$.  We retain these as
discretization systematics; small scan-to-scan oscillations within those bands
are not interpreted as physical nonmonotonicity.

\subsection{Polarimetric response of the amplified sequence}
\label{sec:polresults}

\begin{figure*}[t]
 \centering
 \includegraphics[width=0.99\textwidth]{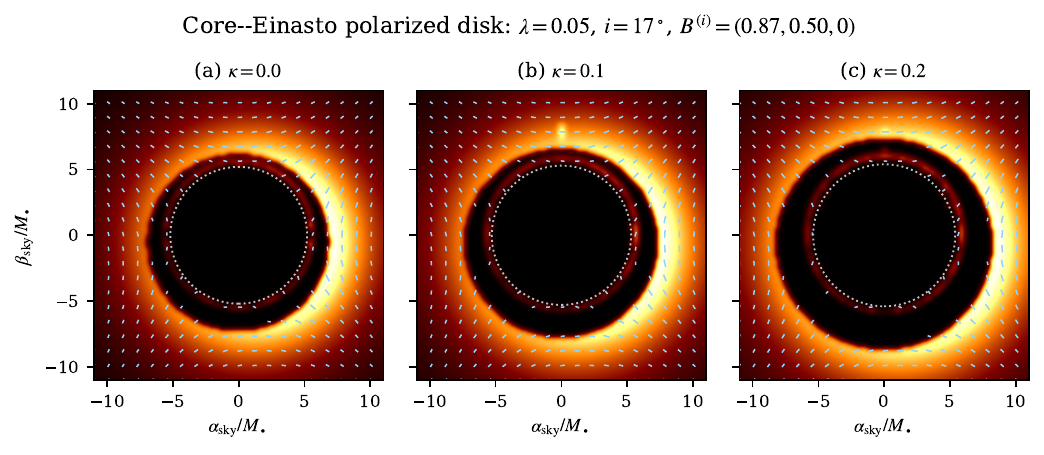}
 \caption{\label{fig:polmaps}
 Total polarized disk images for (a) $\kappa=0$, (b) $\kappa=0.1$, and
 (c) $\kappa=0.2$, with $(\alpha,q,\lambda)=(0.16,0.1,0.05)$,
 $i=17^\circ$, and $B^{(i)}=(0.87,0.50,0)$.  Color gives Stokes $I$
 normalized separately in each panel; blue line segments give the EVPA and
 are shown only where the polarized intensity exceeds $3.5\%$ of its peak.
 The dotted circle is the exact critical curve, not a fitted image boundary.
 These compactnesses make the geometric response visible and are not galactic
 calibrations.}
\end{figure*}

Figure~\ref{fig:polmaps} shows the total Stokes intensity and EVPA ticks.  The
dotted circle is the exact critical curve $b_{\rm sh}$.  Increasing $\kappa$
enlarges both the central lensed region and the disk inner edge.  The bright
side remains set by orbital Doppler weighting, while the polarization field
changes smoothly because the halo modifies the ray bending, emitter velocity,
redshift, and screen transport simultaneously.  Spherical symmetry introduces
no frame dragging; any interpretation of the change as a spin signature would
therefore be incorrect.

Table~\ref{tab:polarization} quantifies the trend on the higher-resolution
maps.  Between $\kappa=0$ and $0.2$, $r_{\rm ISCO}$ increases by $18.3\%$ and
$b_{\rm sh}$ by $3.84\%$.  Within the fixed emission model,
$|m|_{\rm net}$ decreases from $1.143\%$ to $0.778\%$, while
$|\beta_2|$ increases and its phase changes by $-3.07^\circ$.  The phase shift
is larger than the $0.63^\circ$ resolution systematic, whereas the absolute
normalization of $|m|_{\rm net}$ remains sensitive to $p_0$, the magnetic
field, disk boundaries, and neglected Faraday effects.  Higher-order crossings
contribute only $3.2$--$3.9\%$ of the model flux.  They alter the fine
near-critical structure but do not control the global polarization morphology.

\begin{table*}[t]
\centering
\caption{\label{tab:polarization}Geometric and image-domain diagnostics for
the $120^2$ maps.  The phase is quoted relative to the Schwarzschild image to
remove the arbitrary parity and EVPA zero point of the screen.}
\begin{tabular}{cccccccc}
\toprule
$\kappa$ & $r_h/M_\bullet$ & $r_{\rm ISCO}/M_\bullet$ &
$b_{\rm sh}/M_\bullet$ & $|m|_{\rm net}$ [\%] & $|\beta_2|$ &
$\Delta\arg\beta_2$ [deg] & $I_{n\geq2}/I$ [\%]\\
\midrule
0.0 & 2.0000 & 6.0000 & 5.1962 & 1.143 & 0.0450 &  0.00 & 3.20\\
0.1 & 2.0146 & 6.3656 & 5.2934 & 0.963 & 0.0547 & $-0.88$ & 3.76\\
0.2 & 2.0297 & 7.0995 & 5.3959 & 0.778 & 0.0758 & $-3.07$ & 3.86\\
\bottomrule
\end{tabular}
\end{table*}

\begin{figure*}[t]
 \centering
 \includegraphics[width=0.94\textwidth]{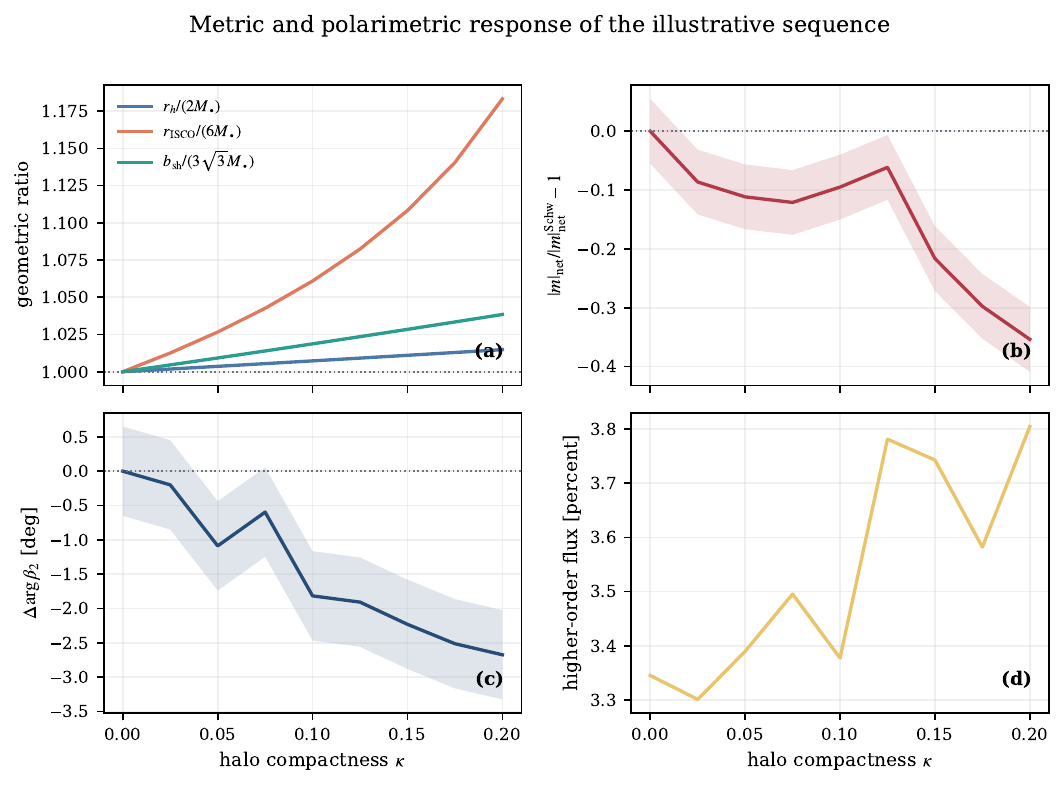}
 \caption{\label{fig:poldiagnostics}
 Response along the amplified $\lambda=0.05$ sequence.  (a) Horizon, ISCO,
 and critical impact parameter normalized to their Schwarzschild values.
 (b) Relative net polarization, (c) change in the phase of $\beta_2$, and
 (d) higher-order flux fraction.  Shading in (b) and (c) records the
 $120^2$--$160^2$ discretization comparison; fluctuations inside it are not
 resolved trends.}
\end{figure*}

The uniform compactness scan in Fig.~\ref{fig:poldiagnostics} separates the
geometric and polarimetric changes.  The horizon and critical curve grow
smoothly, and the ISCO supplies the strongest geometric response.  The overall
decrease of $|m|_{\rm net}$ and rotation of $\arg\beta_2$ are resolved, but the
small wiggles lie within the shaded discretization bands.  The higher-order
flux fraction remains at a few percent throughout.  These results also show
why a single image statistic cannot isolate a halo: changing the disk inner
edge or magnetic topology can readily produce a larger response than the
metric perturbation.

\subsection{Weak-core scaling and EHT relevance}
\label{sec:polrealistic}

The amplified maps should not be confused with the profile extrapolations in
Sec.~\ref{sec_eht}.  Near the center of any smooth core, define
\begin{equation}
 u=8\pi\rho_0M_\bullet^2
 =\kappa\rhoh(0)\lambda^2.                                      \label{eq:polu}
\end{equation}
The lapse has the universal expansion
\begin{equation}
 f(r)=1-\frac{2M_\bullet}{r}-\frac{u}{3}
 \left(\frac{r}{M_\bullet}\right)^2
 +\order\!\left(u\frac{\lambda r}{qM_\bullet},u^2\right).       \label{eq:polcoremetric}
\end{equation}
Consequently, every smooth image functional has a regular response of order
$u$, provided its emitting domain does not cross a caustic or stability
boundary.  From the largest secant slopes in the numerical sequence we take
the deliberately conservative envelopes
\begin{equation}
 \frac{|\Delta |m|_{\rm net}|}{|m|_{\rm net}}<29.4u,
 \qquad |\Delta\arg\beta_2|<185u\ {\rm deg}.                     \label{eq:polenvelope}
\end{equation}
These are not universal coefficients: they summarize this disk and magnetic
field while absorbing its measured numerical scatter.  They are useful as
upper scales because the exact shadow coefficient, by comparison, is
$\delta_{\rm sh}=4.5u+\order(u\lambda/q,u^2)$.

\begin{figure*}[t]
 \centering
 \includegraphics[width=0.94\textwidth]{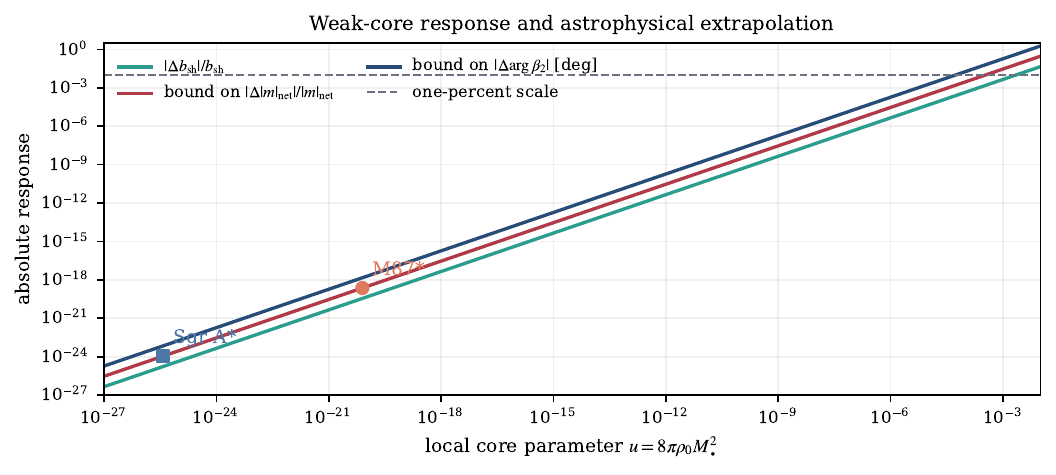}
 \caption{\label{fig:polscale}
 Weak-core shadow response and conservative polarimetric envelopes versus
 $u=8\pi\rho_0M_\bullet^2$.  Markers give the smooth $\mseven$ and $\sgr$
 estimates used in this work.  The dashed line indicates a one-percent scale.
 The straight lines are local $u\rightarrow0$ extrapolations, not direct
 floating-point ray traces at $u\sim10^{-20}$.}
\end{figure*}

The fiducial estimates in Sec.~\ref{sec_eht} give
$u_{\mseven}=7.91\times10^{-21}$ and
$u_{\sgr}=3.84\times10^{-26}$.  Equation~\eqref{eq:polenvelope} then yields
\begin{align}
 \mseven:\quad
 \frac{|\Delta |m|_{\rm net}|}{|m|_{\rm net}}
 &<2.32\times10^{-19},                                           \nonumber\\[-1mm]
 |\Delta\arg\beta_2|&<1.46\times10^{-18}\ \mathrm{deg},           \label{eq:polm87}\\
 \sgr:\quad
 \frac{|\Delta |m|_{\rm net}|}{|m|_{\rm net}}
 &<1.13\times10^{-24},                                           \nonumber\\[-1mm]
 |\Delta\arg\beta_2|&<7.10\times10^{-24}\ \mathrm{deg}.           \label{eq:polsgr}
\end{align}
Figure~\ref{fig:polscale} displays this separation from a one-percent response.
Even the conservative polarimetric envelopes are many orders of magnitude
below observational and plasma-model systematics.

This conclusion is stronger than a failure to fit one polarization statistic.
The EHT finds resolved polarization fractions reaching approximately $15\%$
in the 2017 $\mseven$ image \citep{EHT2021Polarization}, and the 2018--2021
data show lower peak fractions, structural variability, and a reversal of the
EVPA spiral handedness \citep{EHT2025Variability}.  Those order-unity temporal
and plasma effects exceed Eqs.~\eqref{eq:polm87}--\eqref{eq:polsgr} by at
least eighteen orders of magnitude.  Polarized images can test compact matter
distributions or nonvacuum metrics with horizon-scale stress energy, but they
cannot presently constrain a smooth, kiloparsec-scale cEinasto core in the
density-preserving completion.

\section{Discussion}
\label{sec_discussion}

The solution isolates two questions that are often conflated.  The first is
whether an empirical density profile can be embedded exactly in a relativistic
metric.  It can, once an equation-of-state closure is specified.  The second is
whether that closure represents the underlying particle dark matter.  The
The condition $p_r=-\rho$ does not describe a cold collisionless
halo.  We use it to define an effective anisotropic source whose density
matches the desired cEinasto law.  A kinetic Einstein--Vlasov
model or a two-function fluid solution with a microphysically motivated
velocity dispersion would answer the second question and need not share the
near-horizon metric found here.

The smooth extrapolation of a kiloparsec-scale feedback profile to
a few gravitational radii is another simplifying assumption.  Black-hole growth,
stellar scattering, self-annihilation, or self-interactions can create a spike,
crest, plateau, or depleted core.  At leading order, Eq.
\eqref{eq_deltageneral} allows us to replace that inner model using only its
mass inside $3M_\bullet$.  Conversely, the
current EHT bounds in Table \ref{tab_eht} should not be read as competitive
constraints on FIRE-2 halo parameters.  They are consistency bounds on a
smooth, static spherical extrapolation.

Rotation and accretion physics are essential for source-specific EHT
inference.  A Newman--Janis metric is not included because the transformed
stress tensor and its relation to the cEinasto seed would require a separate
consistency analysis.  The static result remains useful as the monopolar
environmental limit and as a base example for future perturbative spinning
solutions.  Likewise, our EHT exercise uses published
shadow-deviation summaries rather than interferometric visibilities.  A
visibility-level study would need a rotating spacetime, radiative transfer, and
joint inference over mass, distance, inclination, and plasma parameters.

\section{Conclusions}
\label{sec_conclusions}

We have derived an analytic, density-preserving
Schwarzschild-gauge geometry for the FIRE-2 cEinasto profile.  The spacetime is
asymptotically flat, its halo mass is finite, and its anisotropic source retains
the prescribed density exactly.  Our closed incomplete-gamma mass function
supports a complete analysis of the horizons, thermodynamics, and null circular
orbits.

Our comparison yields five main findings.  The rotation-curve
prescription replaces a finite core by an inverse-square source cusp, while the
appended Schwarzschild term cancels from the density equation.  A sharp
compactness threshold separates the guaranteed one-horizon domain from a
strong-source phase that can contain three horizons.  Realistic galactic
compactnesses remain far below that threshold.  At fixed halo parameters, the
ordinary first law holds exactly and the weak-halo branch retains negative heat
capacity.  For any weak spherical environment, the leading shadow shift
depends only on the halo mass inside the vacuum photon sphere.  Published EHT
summaries allow central densities far above astrophysical expectations, while
the fiducial Milky-Way shift is negligibly small.

We therefore keep the halo density profile, the metric ansatz, and
the matter model logically distinct.  Present black-hole shadows can restrict
compact inner dark-matter structures, but they do not restrict a smooth
feedback-cored galactic halo at realistic density.

Future research should replace the effective anisotropic source
with kinetic Einstein--Vlasov matter or a two-function fluid with a motivated
velocity dispersion.  We should also extend the static solution to rotation
through a separate field-equation consistency check and combine the resulting
spacetime with radiative transfer and full EHT visibility inference.  Further
work can compare feedback cores with spikes, annihilation plateaus, and
self-interacting inner profiles at fixed black-hole and halo parameters.  Such
comparisons will reveal which strong-field effects follow from the inner mass
distribution and which follow from the assumed stress relations.

\begin{acknowledgments}
A. \"O. and R. P. would like to acknowledge networking support of the COST Action CA21106 - COSMIC WISPers in the Dark Universe: Theory, astrophysics and experiments (CosmicWISPers), the COST Action CA22113 - Fundamental challenges in theoretical physics (THEORY-CHALLENGES), the COST Action CA21136 - Addressing observational tensions in cosmology with systematics and fundamental physics (CosmoVerse), the COST Action CA23130 - Bridging high and low energies in search of quantum gravity (BridgeQG), and the COST Action CA23115 - Relativistic Quantum Information (RQI) funded by COST (European Cooperation in Science and Technology). A. \"O. and R. P. would also like to acknowledge the funding support of SCOAP3. A. \"O. also thanks to EMU, TUBITAK, ULAKBIM (Turkiye)
\end{acknowledgments}

\bibliography{references}

\end{document}